# Molecular beam epitaxy of superconducting FeSe$_x$Te$_{1-x}$ thin films interfaced with magnetic topological insulators

Yuki Sato[1,*,†], Soma Nagahama[2,*], Ilya Belopolski[1], Ryutaro Yoshimi[1],

Minoru Kawamura[1], Atsushi Tsukazaki[3], Naoya Kanazawa[4],

Kei S. Takahashi[1], Masashi Kawasaki[1,2], and Yoshinori Tokura[1,2,5]

[1]*RIKEN Center for Emergent Matter Science (CEMS), Wako 351-0198, Japan*

[2]*Department of Applied Physics and Quantum-phase Electronics Center (QPEC),*

*University of Tokyo, Tokyo 113-8656, Japan*

[3]*Institute for Materials Research (IMR), Tohoku University, Sendai 980-8577, Japan*

[4]*Institute of Industrial Science, University of Tokyo, Tokyo 153-8505, Japan*

[5]*Tokyo College, University of Tokyo, Tokyo 113-8656, Japan*

*\*These authors contributed equally to this work.*

†yuki.sato.yj@riken.jp

Engineering heterostructures with various types of quantum materials can provide an intriguing playground for studying exotic physics induced by the proximity effect. Here, we report the successful synthesis of iron-based superconductor FeSe$_x$Te$_{1-x}$ (FST) thin films across the entire composition range of $0 \leq x \leq 1$ and its heterostructure with a magnetic topological insulator by using molecular beam epitaxy. Superconductivity is observed in the FST films with an optimal superconducting transition temperature $T_c$ ~ 12 K at around $x = 0.1$. We found that superconductivity survives in the very Te-rich films ($x \leq 0.05$), showing stark contrast to bulk crystals with suppression of superconductivity due to an appearance of bicollinear antiferromagnetism accompanied by a monoclinic structural transition. By examining thickness $t$ dependence of magnetic susceptibility and electrical transport properties, we observed a trend where anomalies associated with the first order structural transition broaden in films with below $t$ ~ 100 nm. We infer this observation suggests a suppression of the structural instability near substrates. Furthermore, we fabricated an all chalcogenide-based heterointerface between FST and a magnetic topological insulator (Cr,Bi,Sb)$_2$Te$_3$ for the first time, observing both superconductivity and a large anomalous Hall conductivity. The anomalous Hall conductivity increases with decreasing temperature, approaching the quantized value of $e^2/h$ down to the measurable minimum temperature at $T_c$. The result suggests coexistence of magnetic and superconducting gaps at low temperatures opening at the top and bottom surfaces, respectively. Our novel magnetic topological insulator/superconductor heterostructure could be an ideal platform to explore chiral Majorana edge mode.

# I.     INTRODUCTION

An interface between different types of quantum materials can host exotic physics originating from non-trivial topology, proximity effect, and broken inversion symmetry. Engineering heterointerfaces may facilitate the establishment of a topological superconductor as a platform for detecting and manipulating Majorana quasiparticles. So far, it has been proposed that Majorana zero mode can be realized at the core of a superconducting vortex in bulk superconductors with a superconducting gap possessing non-trivial topology [1, 2]. Another proposed approach is to introduce the superconducting proximity effect on the surface states of a topological insulator (TI) [3]. Advances in thin film growth techniques have enabled us to fabricate them with great controllability in their electronic structures. For example, it has been established that one can materialize a quantum anomalous Hall (QAH) insulator by substituting magnetic ions in prototypical chalcogenide TI thin films of $(Bi,Sb)_2Te_3$ and electrostatically tuning their Fermi level within the mass gap [4, 5]. One successful example is $(Cr,Bi,Sb)_2Te_3$ (CBST), where inclusion of Cr introduces ferromagnetic ordering below the Curie temperature $T_{Curie}$ and opens a magnetic gap in the Dirac cone. When such QAH insulator is proximitized to a superconductor, non-trivial superconducting phase with broken time reversal symmetry is expected, leading to the emergence of chiral Majorana edge mode [6]. To achieve proximity-induced topological superconductivity, a high-quality interface would be required so that wave functions of two different materials can be hybridized with each other.

One of the promising candidates for a topological superconductor is iron-based superconductor $FeSe_xTe_{1-x}$ (FST). FST has the simplest crystal structures among the iron-based superconductor materials family, which only consists of quasi-two-dimensional stacks of iron-chalcogen sheets. While FeSe ($x = 1$) is a trivial superconductor, isovalent Te substitution into Se can enhance spin-orbit coupling and it is claimed that a topologically non-trivial phase is realized at $x = 0.5$, pointed out by first principles calculation and angle-resolved photoemission spectroscopy (ARPES) study [7]. As the topological surface state is expected to be directly proximitized to its bulk superconductivity below the superconducting transition temperature $T_c$, the surface of FST can be regarded as a two-dimensional $p$-wave superconductor. Indeed, the topological surface state and isotropic gap at the Dirac cone are observed in ARPES study [8], providing evidence for topological superconductivity. Furthermore, scanning tunneling microscopy (STM) reveals zero-bias peak in tunneling conductance [9–11], suggesting the presence of a Majorana zero mode in vortex cores. Another remarkable feature of FST is its excellent affinity to chalcogenide TI thin films. It is reported that TI thin film $Bi_2Te_3$ (BT), thanks to its van der Waals stacking nature, can be epitaxially grown on top of the parent compound FeTe ($x$ = 0) using the molecular beam epitaxy (MBE) technique [12]. Additionally, nonreciprocal transport is observed in this all chalcogenide-based heterostructure, suggesting the crucial roles of proximity effect and broken inversion symmetry [13].

However, the tunneling conductance peaks observed in the STM studies do not always appear at zero-energy, suggesting that a trivial origin can also play a role [14]. One plausible obstacle is the inevitable inhomogeneity in the sample near the half composition $x$ = 0.5, where FST phase-separates into Se-rich and Te-rich clusters. It is proposed that the inhomogeneity causes topological phase separation, and signatures of the non-trivial phase can be found only in the Te-rich region [10, 15]. Although such inhomogeneity might be suppressed in FST near the end compound FeTe, it enters a bicollinear antiferromagnetic (AF) state below $x \sim 0.05$ accompanied by a monoclinic structural transition at about 70 K [16], which readily destroys superconductivity [17, 18]. Therefore, a Te-rich superconducting FST with negligible inhomogeneity and strong spin-orbit coupling, which can further stabilize the topological superconducting phase, is in high demand but still lacking.

In this paper, we demonstrate successful growth of FST thin films on CdTe(100) substrate covering the entire composition of $0 \leq x \leq 1$ by using MBE technique. We found that all FST films with a thickness $t$ = 40 nm exhibit superconductivity, with an optimal $T_c \sim 12$ K observed at around $x$ = 0.1. Surprisingly, superconductivity survives even in films with a very rich Te content, showing a stark contrast to the behavior observed in bulk crystals. By tracking magnetic and transport properties with varying $t$, we found that anomalies associated with the structural transition are weak in films thinner than about 100 nm. This observation may suggest that substrate effect suppresses the structural transition and modifies the electronic structure near the interface. We further engineered an all chalcogenide-based heterointerface between the magnetic topological insulator CBST and the superconducting FST. The CBST/FST heterostructure exhibits both superconductivity and large anomalous Hall conductivity that approaches the quantized value of $e^2/h$ at low temperatures. This exotic heterostructure, therefore, fulfills the prerequisite for the emergence of chiral Majorana edge mode. Our work thus takes an important step toward establishing a topological superconductor as a materials platform to hunt Majorana excitations.

## II. METHOD

All thin films presented in this work were grown by MBE technique. The base pressure of the growth chamber was below $1 \times 10^{-6}$ Pa. High-purity sources of Fe (99.999 %), Se (99.9999 %), Te (99.999 %), Cr (99.99 %), Bi (99.9999 %), and Sb (99.9999 %) were evaporated by heating individual Knudsen-cells and by stabilizing their temperatures to obtain stable and well-controlled equivalent beam fluxes. The amounts of the supplied fluxes were monitored by equivalent pressure of the beam flux monitor. The flux ratio for FST is Fe : Se : Te = 1 : $10x_n$ : $10(1-x_n)$, where the beam equivalent pressure of Fe is $3.5 \times 10^{-6}$ Pa, and the nominal flux ratio $x_n$ varies from 0 to 1 to control the Se content in the FST films. The flux ratio for CBST is Cr : Bi : Sb : Te = 1 : 19 : 30 : 175, where the beam equivalent pressure of Cr is $2.0 \times 10^{-7}$ Pa. Although the actual composition $x$ in the films is found to be regulated by $x_n$, we calibrated $x$ using the inductively coupled plasma mass spectrometry analysis (See details in Supplemental Materials [19]). Insulating non-doped CdTe(100) substrates with a high resistivity exceeding $1 \times 10^7$ $\Omega$cm were etched with bromine-methanol (bromine 0.01%) for 5 minutes before the deposition [20]. Both FST and CBST were fabricated at 240°C. After the growth, we kept supplying Te flux during cooling samples down to around 100°C for the FST films with $x_n \leq 0.6$. This Te-annealing process turns to be of great importance to improve normal state metallicity and to stabilize superconductivity [17, 19, 21]. As we found the Te-annealing can also cause severe phase separation for samples with higher Se concentration ($x_n > 0.6$), we annealed those Se-rich FST films in a vacuum at 360°C for 30 minutes. We found the actual Se content $x$ is hardly affected by these annealing processes [19]. For the growth of heterostructures of the Te-rich FST and the magnetic topological insulator CBST, we deposited CBST after the growth of FST without Te-annealing. $AlO_x$ capping layer was deposited on CBST/FST heterostructures by atomic layer deposition immediately after taking them out of the MBE chamber. X-ray diffraction (XRD) measurements were carried out using SmartLab (RIGAKU) with Cu K$\alpha$ X-ray. Electrical transport and magnetic susceptibility measurements were performed by Physical Property Measurement System and Magnetic Property Measurement System (Quantum design), respectively.

### III. EXPERIMENTAL RESULTS AND DISCUSSIONS

#### A. Growth, characterization, and transport properties of FeSe$_x$Te$_{1-x}$ thin films

We demonstrate that FST thin films can be epitaxially grown on CdTe(100) substrate, covering the entire substitution range ($0 \le x \le 1$). Figure 1(a) shows a cross-sectional high-angle annular dark-field scanning transmission electron microscopy (HAADF-STEM) image of an FST thin film ($x = 0.1$) after the Te-annealing procedure. There can be seen a sharp interface between the FST film and the CdTe substrate. The energy dispersive X-ray spectroscopy (EDX) analysis shows abrupt changes of comprising elements at the surface (Fig. 1(b)). We note that both Se and Te are distributed homogeneously throughout the whole FST region without phase separation, which often occurs in solid-state reaction syntheses [22]. Besides, in the capping layer, which is formed during the Te-annealing process, we found the atomic ratio of Fe : Te $\sim 1 : 2$, suggesting formation of FeTe$_2$ phase. In the Supplemental Materials, we discuss possible formation of insulating marcasite FeTe$_2$ on top of FST [19], which however should hardly affect the observed transport data of FST. X-ray diffraction (XRD) patterns are summarized in Fig. 1(c) for FST thin films ($t = 40$ nm) with different nominal compositions: $x = 0$, 0.38, 0.6, 0.72, and 1. Sharp peaks from FST(00$n$) are clearly observed for all the samples, demonstrating successful growth of homogeneous films without phase separation. These peaks are systematically shifted to higher angles as $x$ increases. Corresponding lattice constant $c$ is calculated from the peak position by using Bragg's law and is plotted as a function of $x$ in Fig. 1(d). We found the $c$-axis length of the films agree well with those of single crystals with a nearly linear relation, suggesting that Vegard's law holds throughout the entire substitution range. Apart from the FST(00$n$) peaks, all the other features found in the XRD data can be attributed to CdTe substrate, which contributes to sharp (00$n$) peaks and tiny NaCl type impurity phase labelled by down-pointing triangles and asterisk, respectively.

We investigated electrical transport properties of the FST thin films. Resistivity-temperature, ($\rho_{xx}$-$T$) curves for the FST films ($0 \le x \le 1$) with $t = 40$ nm are presented in Figs. 2(a) and 2(b). FeTe ($x = 0$) shows insulating behavior at high temperatures, peaks around 70 K, and metallic behavior at low temperatures. As $x$ increases, this insulator-metal crossover temperature, denoted by arrows in Fig. 2(a), tends to shift to higher temperatures. Such an insulator-metal crossover with increasing $x$ is also reported in electrical transport [18] and ARPES studies [23] on bulk single crystals, where the Te-annealing procedure was also employed. Without the Te-annealing procedure, the metallic behavior was absent, and the insulating behavior persisted down to $T_c$ [19]. These results suggest that the present FST thin films in Figs. 2(a) and 2(b) were sufficiently Te-annealed. Although the overall observed transport property of the thin films is similar to that of bulk crystals, there are several distinct features of the thin films. First, we observe superconducting transitions (resistivity drops) for the entire composition range, including even at $x = 0$ (FeTe) (Fig. 2(b)). This observation contrasts with the report that superconductivity in the bulk single crystal is strongly suppressed below $x \sim 0.05$ [17, 18], where superconductivity competes with the AF ordering and/or the AF order induced monoclinic structural distortion. Second, as shown in Fig. 2(c), the superconducting dome for our thin films peaks around $x \sim 0.1$, reaching an optimal $T_c \sim 12$ K, which is quantitatively different from that of both bulk single crystals and other thin films grown on different substrates [24-26]. Third, the resistivity anomaly related to the structural phase transition at $T_s$ cannot be clearly seen in our thin films, while bulk single crystals of the end compounds FeSe and FeTe are known to show kink and jump anomalies at $T_s \sim 90$ K and 70 K, respectively. We will discuss this point in the following in more detail.

### B. Possible suppression of the first-order structural transition in thin film FeSe$_x$Te$_{1-x}$

When materials are grown on substrates in the form of epitaxial thin films, their physical properties can be greatly affected by strain, charge transfer, and phonon coupling from the substrates [27, 28]. This substrate effect appears to be of great importance especially for FST thin films, where multiple degrees of instability such as superconductivity, magnetism, and orbital ordering compete with each other. This is likely the reason why the phase diagrams of FST are dramatically different depending on how the samples are grown, as shown in Fig. 2(c). Indeed, it has been pointed out that changes in lattice constants and bond angles can significantly influence $T_c$ in both FeSe and FeTe thin films [29, 30]. Furthermore, it is claimed that the substrate can substantially suppress the lattice structural transition in thin films of FeSe [31], which still holds anisotropy in the electronic system, suggesting an electronically driven nematic state [32].

To check whether the observed transport properties are inherent features of FST thin films grown on CdTe substrate, we examine $t$ dependence of transport properties to see how the thin film properties recover to those of bulk single crystals as $t$ is increased. In Fig. 3(a), we present magnetic susceptibility $\chi$ of three FeTe thin films with $t = 40$, 250, and 1000 nm as a function of $T$. The thickest sample ($t = 1000$ nm) shows an abrupt drop below around 60 K, which corresponds to the Néel temperature $T_N$. However, as $t$ decreases, the sharpness of the AF transition gets broader, and no significant drop in $\chi$ was observed at $T_N$ for the thinnest sample ($t = 40$ nm). This broadening of the transition can be also seen in Hall coefficient $R_H$, as shown in Fig. 3(b). In the $t = 1000$ nm sample, $R_H$ shows sharp reduction below the $T_N$, and changes its sign from positive to negative. An even more sudden reduction of $R_H$ and a sign change were also reported in the bulk single crystals [17], which is considered to be a consequence of Fermi surface reconstruction driven by bicollinear AF ordering accompanied by the lattice structural transition to the monoclinic form. As $t$ decreases, however, $R_H$ becomes larger in magnitude and the anomaly near $T_N$ gets broader, showing no sign change down to 2 K in the $t$ = 40 nm film. Similar effect can be found in $\rho_{xx}$-$T$ curves of FeTe thin films as shown in Fig. 3(c). While the thicker samples show a more pronounced drop of $\rho_{xx}$ below the $T_N$, the thinner sample shows a broader feature with relatively large resistivity. Because the sudden changes at the $T_N$ are likely due to the first-order phase transition nature of the structural instability, the broadening of $\chi$ and $R_H$ in the thinner samples implies that the structural instability is somewhat suppressed in the FeTe thin films. In addition to such suppressed structural transition, possible influence from the substrate can be also seen even at higher temperatures. To see the $t$-driven evolution of transport properties, we show $t$ dependence of conductivity $\sigma_{xx}$ and $1/eR_H$, reflecting carrier density in the case of a single-band model, at $T$ = 150 K well above the $T_N$ in Fig. 3(d). While both the quantities hold comparable values in thick samples with $t \geq 200$ nm, they exhibit sudden drops below $t \sim 100$ nm. These results imply that the electronic structure near the substrate might be different from that of the rest of the film, and that such region expands from the interface up to 100 nm or less, as schematically illustrated in the inset of Fig. 3(d). If such region is superconducting, it is expected that superconductivity will be observed irrespective of $t$, even if the remaining parts are metallic. Indeed, we observed onset of superconductivity in all the three FeTe films we measured, as shown in Fig. 3(c) (See also Supplemental Materials [19]). It is interesting to know whether the FeTe thin films possess nearly perfect compensation of electron and hole, or total carrier concentration is progressively reduced. However, it is difficult to estimate carrier concentration and mobility for both electron and hole separately, because Hall resistivity shows nearly linear in $H$-dependence [19] up to 14 T. Further study is needed to clarify the fermiology in FeTe and its systematic evolution with $x$ in these FST thin films.

We make a comparison between the phase diagrams for bulk single crystals and our MBE grown FST films with $t$ = 40 nm shown in Figs. 3(e) and 3(f), respectively, focusing on the Te-rich region ($0 \leq x \leq 0.2$). In the bulk crystals, the AF order accompanied by the monoclinic lattice distortion takes place below $x \sim 0.05$, and superconductivity is strongly suppressed in this region. In our MBE films, on the other hand, superconductivity survives even in the vicinity of $x$ = 0, where superconductivity has never been observed in bulk. We note that $T_c$ here is determined by the zero-resistivity temperature, and that the drop in $\rho_{xx}$ is observed even at $x$ = 0. More detailed transport data in this Te-rich region are available in the Supplemental Materials [19]. To understand the difference of the phase diagrams, it is important to clarify whether the AF phase coexists with superconductivity.

Recent spin-resolved STM studies revealed that the AF order holds in a few-layer FeTe films grown on different substrates [33, 34]. The substrate effects appear to prevent the lattice structural transition in the FeTe while they might keep AF fluctuation or ordering intact. Consequently, the second-order AF transition could lead to the gradual changes as observed in the $\rho_{xx}$ and $R_H$ below $T \sim$ 70 K in the $t$ = 40 nm sample. The suppressed lattice structural transition suggests a potential difference in the Fermi surface topology between the thinner FeTe films and bulk single crystals, which could serve as a pivotal factor in stabilizing superconductivity in the Te-rich FST. However, one might interpret the gradual changes as observed in the the $\rho_{xx}$ and $R_H$ as an incoherent-coherent crossover [18], or more specifically, Kondo-type band renormalization [35]. We need further investigation on the presence of long-range or short-range magnetic order in our FST films and its influence on the observed superconductivity.

### C.  Growth, characterization, and transport properties of (Cr,Bi,Sb)$_2$Te$_3$/FeSe$_x$Te$_{1-x}$ heterostructures

We discuss the heterointerfaces between the magnetic TI CBST and the superconductor FST. It is reported that BT, the mother compound of CBST, can be grown on FeTe using MBE technique, even though they have different lateral rotational symmetries of $C_4$ (FeTe) and $C_3$ (BT) [12, 13] (dubbed hybrid epitaxy [26]). This unusual situation enables us to fabricate heterostructures even in the opposite stack order, i.e., FST/BT [36, 37]. Interestingly, the heterostructure (BT/FT) between those non-superconducting materials can host interfacial superconductivity. Furthermore, nonreciprocal transport is observed in the BT/FT heterostructure, suggesting a strong coupling between the topological surface state and the superconducting proximity effect [13].

Figures 4(a) and 4(b) show HAADF-STEM picture and EDX elemental mappings for a CBST/FST heterostructure. We observed sharp interface between FST and CBST. Each element composing CBST and FST is homogeneously distributed and confined to the respective intended layers, demonstrating a high-quality heterointerface. In addition, we observed significant XRD intensities from both FST and CBST, as shown in Fig. 4(c). The observed peak positions well agree with those of thin films of CBST and FST as indicated by blue and red down-triangles, respectively, suggesting a nearly distortion-free heterointerface. We also examined surface topographies taken by atomic force microscopy for as-grown FST and CBST/FST films (Fig. 4(d)). The topography of FST is covered by small and squared patterns, whose edge is aligned to CdTe(100) direction. After the deposition of CBST, the above-mentioned structure disappears and larger terrace and step like structure covering the whole scanned area shows up, which is reminiscent of that of MBE-grown CBST thin films [38].

We investigate transport properties of the CBST/FST heterostructure. Because the presence of Se in CBST is detrimental to the magnetic order necessary for the quantum anomalous Hall effect, it is highly desirable to develop an all-Te heterostructure with minimal Se level. For this reason, it is essential to pursue Te-rich superconducting FST films in combination with ferromagnetic CBST approaching quantization. In Fig. 5(a), we present longitudinal sheet resistivity $R_{xx}$ of $(Cr_{0.02}Bi_{0.38}Sb_{0.6})_2Te_3/FeSe_{0.1}Te_{0.9}$ with different $t$ of FST as a function of $T$. The nominal composition used for the CBST layer is similar to the optimal conditions for observing the QAH effect in CBST single layers [5, 39]. In a sample with $t$ = 45 nm, superconductivity is observed at $T_c \sim 12$ K, which is as high as that of FST single layer with $x$ = 0.1. This indicates that at $t$ = 45 nm, superconductivity in FST is hardly affected by CBST. As $t$ decreases, however, we observe a reduction in $T_c$, and the system finally becomes non-superconducting at $t$ = 4 nm. In the case of non-magnetic heterostructure BT/FST, superconductivity is observed even at $t$ = 4 nm [19]. Taking this fact into account, the absence of superconductivity in the $t$ = 4 nm CBST/FST suggests that superconductivity is partly suppressed due to the magnetic proximity effect. In Fig. 5(b), we plot sheet conductance $\sigma_{xx} = \rho_{xx}/(\rho_{xx}^2 + \rho_{yx}^2)$ of the CBST/FST films in the normal state ($T$ = 20 K) as a function of $t$. We observe an almost linear relationship between $\sigma_{xx}$ and $t$ with a nearly zero intercept, indicating that the longitudinal conductivity is mainly determined by FST. Therefore, CBST remains highly insulating, indicating that the Fermi energy in CBST is close to the charge neutral point, which is a prerequisite for the QAH effect. The negative intercept of $\sigma_{xx}$ may indicate that the transport for the $t$ = 4 nm sample is severely affected by the magnetic proximity effect from CBST, which enhances magnetic scattering showing very insulating behavior. Importantly, we observe the anomalous Hall effect in the CBST/FST heterostructure, whose amplitude at zero field rapidly grows with lowering temperature (Fig. 5(c)). The positive slope of $\sigma_{xy} = \rho_{yx}/(\rho_{xx}^2 + \rho_{yx}^2)$ against the magnetic field $H$ in the paramagnetic state around 70 K, where the anomalous Hall signal disappears, is comparable to that of FST [19], indicating that the ordinary Hall effect of CBST is negligibly small. We then plot the anomalous Hall conductivity $^A\sigma_{xy}$, obtained by extrapolation from high field, against $T$ (Fig. 5(d)). $^A\sigma_{xy}$ starts to increase at $T_{Curie} \sim$ 65 K and reaches as high as $0.8e^2/h$ at 10 K, above the superconducting transition. Although observing the QAH effect in CBST requires very low temperature, typically well below 1 K [5], we cannot measure the Hall conductance of CBST/FST at lower temperatures because the superconducting FST completely shunts Hall voltage below $T_c$. However, the reasonably large $^A\sigma_{xy} \sim 0.8e^2/h$ at 10 K and the $T_{Curie}$ of about 65 K are comparable to those for CBST thin films, which show the well-defined QAH effect at temperatures below 1 K [39]. Therefore, it is natural to consider that the CBST/FST heterostructure has a well-developed exchange gap on the CBST side and a superconducting gap on the FST side at temperatures below $T_c$.

The observed large anomalous Hall conductivity approaching the QAH state and superconductivity in the CBST/FST heterostructure fulfills the prerequisites for the emergence of chiral Majorana edge modes. We note that this exotic quantum Hall insulator/ topological superconductor heterostructure may be utilized to detect much concrete evidence of multiple chiral Majorana edge modes [40]. Recently, it was predicted that chiral Majorana edge modes are also detectable through local optical conductivity [41]. This work therefore may trigger further experimental attempts to investigate Majorana excitations in topological superconductors.

## IV. CONCLUSION

To summarize, we demonstrate the successful fabrication of homogeneous FST thin films by using MBE method throughout the entire composition of $0 \leq x \leq 1$. The FST films show the single superconducting dome which peaks at ~ 12 K around the optimal composition of $x = 0.1$ in the $T_c$-$x$ phase diagram. Remarkably, superconductivity can survive in the very Te-rich composition of $x \leq 0.05$, whereas it is completely suppressed in bulk single crystals. We attribute this difference in phase diagrams to possible substrate effects that substantially suppress the first-order structural transition and greatly modifies the electronic structure. The extremely Te-rich superconducting FST may not suffer from the inhomogeneity problem, making it an ideal candidate for studying Majorana zero modes by STM. Additionally, we demonstrate the fabrication of a novel heterointerface between magnetic TI (CBST) and superconductor (FST), both based on tellurides with large spin-orbit coupling and possibly small carrier densities. This heterostructure exhibits a large anomalous Hall conductivity in the normal state and superconductivity below $T_c$. Our results, therefore, present an important step toward establishing a concrete materials platform to explore Majorana excitations in solids.

During the peer review process, we became aware of a related study on CBST/FeTe heterostructures [42]. Our CBST/FST heterostructure has the same composition except for the inclusion of Se into FeTe. The advantage of Se doping is that it can provide a more robust bulk-origin superconducting proximity effect to the magnetic topological insulator, while maintaining a large magnetic gap. Indeed, the CBST/FST heterostructure allows us to incorporate high Cr concentration in CBST, thereby maintaining both robust superconductivity and a large anomalous Hall conductivity approaching $e^2/h$.

## ACKNOWLEDGMENTS

We are thankful to N. Nagaosa, X. Yu, T. Shibauchi, and C.-K. Chiu for fruitful discussions. This work was supported by JSPS KAKENHI Grants (No. 22K13988, No. 22K18965, No. 23H04017, No. 23H05431, and No. 23H05462), JST FOREST (Grant No. JPMJFR2038), JST CREST (Grant No. JPMJCR1874 and No. JPMJCR23O3), Mitsubishi Foundation, and the special fund of Institute of Industrial Science, The University of Tokyo.

## REFERENCES

[1] M. Sato and Y. Ando, Topological superconductors: a review, Reports on Progress in Physics **80**, 076501 (2017).

[2] A. P. Schnyder and P. M. Brydon, Topological surface states in nodal superconductors, Journal of Physics: Condensed Matter **27**, 243201 (2015).

[3] L. Fu and C. L. Kane, Superconducting proximity effect and Majorana fermions at the surface of a topological insulator, Physical Review Letters **100**, 096407 (2008).

[4] C.-Z. Chang, J. Zhang, X. Feng, J. Shen, Z. Zhang, M. Guo, K. Li, Y. Ou, P. Wei, L.-L. Wang, *et al.*, Experimental observation of the quantum anomalous hall effect in a magnetic topological insulator, Science **340**, 167 (2013).

[5] Y. Tokura, K. Yasuda, and A. Tsukazaki, Magnetic topological insulators, Nature Reviews Physics **1**, 126 (2019).

[6] J. Wang, Q. Zhou, B. Lian, and S.-C. Zhang, Chiral topological superconductor and half-integer conductance plateau from quantum anomalous hall plateau transition, Physical Review B **92**, 064520 (2015).

[7] Z. Wang, P. Zhang, G. Xu, L. Zeng, H. Miao, X. Xu, T. Qian, H. Weng, P. Richard, A. V. Fedorov, *et al.*, Topological nature of the FeSe$_{0.5}$Te$_{0.5}$ superconductor, Physical Review B **92**, 115119 (2015).

[8] P. Zhang, K. Yaji, T. Hashimoto, Y. Ota, T. Kondo, K. Okazaki, Z. Wang, J. Wen, G. Gu, H. Ding, *et al.*, Observation of topological superconductivity on the surface of an iron-based superconductor, Science **360**, 182 (2018).

[9] D. Wang, L. Kong, P. Fan, H. Chen, S. Zhu, W. Liu, L. Cao, Y. Sun, S. Du, J. Schneeloch, *et al.*, Evidence for Majorana bound states in an iron-based superconductor, Science **362**, 333 (2018).

[10] L. Kong, S. Zhu, M. Papaj, H. Chen, L. Cao, H. Isobe, Y. Xing, W. Liu, D. Wang, P. Fan, *et al.*, Half-integer level shift of vortex bound states in an iron-based superconductor, Nature Physics **15**, 1181 (2019).

[11] T. Machida, Y. Sun, S. Pyon, S. Takeda, Y. Kohsaka, T. Hanaguri, T. Sasagawa, and T. Tamegai, Zero-energy vortex bound state in the superconducting topological surface state of Fe(Se, Te), Nature materials **18**, 811(2019).

[12] Q. L. He, H. Liu, M. He, Y. H. Lai, H. He, G. Wang, K. T. Law, R. Lortz, J. Wang, and I. K. Sou, Two-dimensional superconductivity at the interface of a Bi$_2$Te$_3$/FeTe heterostructure, Nature communications **5**, 4247 (2014).

[13] K. Yasuda, H. Yasuda, T. Liang, R. Yoshimi, A. Tsukazaki, K. S. Takahashi, N. Nagaosa, M. Kawasaki, and Y. Tokura, Nonreciprocal charge transport at topological insulator/superconductor interface, Nature communications **10**, 2734 (2019).

[14] T. Machida and T. Hanaguri, Searching for Majorana quasiparticles at vortex cores in iron-based superconductors, Progress of Theoretical and Experimental Physics, ptad084 (2023).

[15] Y. Li, N. Zaki, V. O. Garlea, A. T. Savici, D. Fobes, Z. Xu, F. Camino, C. Petrovic, G. Gu, P. D. Johnson, *et al.*, Electronic properties of the bulk and surface states of $Fe_{1+y}Te_{1-x}Se_x$, Nature Materials **20**, 1221 (2021).

[16] W. Bao, Y. Qiu, Q. Huang, M. Green, P. Zajdel, M. Fitzsimmons, M. Zhernenkov, S. Chang, M. Fang, B. Qian, *et al.*, Tunable ($\delta\pi$, $\delta\pi$)-type antiferromagnetic order in $\alpha$-Fe(Te,Se) superconductors, Physical Review Letters **102**, 247001 (2009).

[17] Y. Sun, T. Yamada, S. Pyon, and T. Tamegai, Influence of interstitial Fe to the phase diagram of $Fe_{1+y}Te_{1-x}Se_x$ single crystals, Scientific reports **6**, 32290 (2016).

[18] T. Otsuka, S. Hagisawa, Y. Koshika, S. Adachi, T. Usui, N. Sasaki, S. Sasaki, S. Yamaguchi, Y. Nakanishi, M. Yoshizawa, *et al.*, Incoherent-coherent crossover and the pseudogap in Te-annealed superconducting $Fe_{1+y}Te_{1-x}Se_x$ revealed by magnetotransport measurements, Physical Review B **99**, 184505 (2019).

[19] See Supplemental Materials.

[20] M. Goyal, L. Galletti, S. S-Rezair, T. Schumann, D. A. Kealhofer, S. Stemmer, Thickness dependence of the quantum Hall effect in films of the three-dimensional Dirac semimetal $Cd_3As_2$, APL Matter **6**, 026105 (2018).

[21] Y. Sun, Z. Shi, and T. Tamegai, Review of annealing effects and superconductivity in $Fe_{1+y}Te_{1-x}Se_x$ superconductors, Superconductor Science and Technology **32**, 103001 (2019).

[22] M. H. Fang, H. M. Pham, B. Qian, T. J. Liu, E. K. Vehstedt, Y. Liu, L. Spinu, and Z. Q. Mao, Superconductivity close to magnetic instability in $Fe(Se_{1-x}Te_x)_{0.82}$, Physical Review B **78**, 224503 (2008).

[23] J. Huang, R. Yu, Z. Xu, J.-X. Zhu, J. S. Oh, Q. Jiang, M. Wang, H. Wu, T. Chen, J. D. Denlinger, *et al.*, Correlation-driven electronic reconstruction in $FeTe_{1-x}Se_x$, Communications Physics **5**, 29 (2022).

[24] K. Mukasa, K. Matsuura, M. Qiu, M. Saito, Y. Sugimura, K. Ishida, M. Otani, Y. Onishi, Y. Mizukami, K. Hashimoto, *et al.*, High-pressure phase diagrams of $FeSe_{1-x}Te_x$: correlation between suppressed nematicity and enhanced superconductivity, Nature communications **12**, 381 (2021).

[25] Y. Imai, Y. Sawada, F. Nabeshima, and A. Maeda, Suppression of phase separation and giant enhancement of superconducting transition temperature in $FeSe_{1-x}Te_x$ thin films, Proceedings of the National Academy of Sciences **112**, 1937 (2015).

[26] X. Yao, M. Brahlek, H. T. Yi, D. Jain, A. R. Mazza, M.-G. Han, and S. Oh, Hybrid symmetry epitaxy of the superconducting Fe(Te, Se) film on a topological insulator, Nano Letters **21**, 6518 (2021).

[27] D. Huang and J. E. Hoffman, Monolayer FeSe on $SrTiO_3$, Annual Review of Condensed Matter Physics **8**, 311 (2017).

[28] G. Phan, K. Nakayama, K. Sugawara, T. Sato, T. Uata, Y. Tanabe, K. Tanigaki, F. Nabeshima, Y. Imai, A. Maeda, *et al.*, Effects of strain on the electronic structure, superconductivity, and nematicity in FeSe studied by angle-resolved photoemission spectroscopy, Physical Review B **95**, 224507 (2017).

[29] F. Nabeshima, M. Kawai, T. Ishikawa, N. Shikama, and A. Maeda, Systematic study on transport properties if FeSe thin films with various degrees of strain, Japanese Journal of Applied Physics **57**, 120314 (2018).

[30] Y. Han, W. Y. Li, L. X. Cao, X. Y. Wang, B. Xu, B. R. Zhao, Y. Q. Guo, and J. L. Yang, Superconductivity in iron telluride thin films under tensile stress, Physical Review Letters **104**, 017003 (2010).

[31] D. Huang, T. A. Webb, S. Fang, C.-L. Song, C.-Z. Chang, J. S. Moodera, E. Kaxiras, and J. E. Hoffman, Bounds on nanoscale nematicity in single-layer $FeSe/SrTiO_3$, Physical Review B **93**, 125129 (2016).

[32] Y. Kubota, F. Nabeshima, K. Nakayama, H. Ohsumi, Y. Tanaka, K. Tamasaku, T. Suzuki, K. Okazaki, T. Sato, A. Maeda, *et al.*, Pure nematic state in the iron-based superconductor FeSe, Physical Review B **108**, L100501(2023).

[33] T. Hänke, U. R. Singh, L. Cornils, S. Manna, A. Kamlapure, M. Bremholm, E. M. J. Hedegaard, B. B. Iversen, P. Hofmann, J. Hu, *et al.*, Reorientation of the diagonal double-stripe spin structure at $Fe_{1+y}Te$ bulk and thin film surfaces, Nature Communications **8**, 13939 (2017).

[34] S. Sharma, H. Li, Z. Ren, W. A. Castro, and I. Zeljkovic, Nanoscale visualization of the thermally driven evolution of antiferromagnetic domains in FeTe thin films, Physical Review Materials **7**, 074401 (2023).

[35] Y. Kim, M.-S. Kim, D. Kim, M. Kim, M. Kim, C.-M. Cheng, J. Choi, S. Jung, D. Lu, J. H. Kim, *et al.*, Kondo interaction in FeTe and its potential role in the magnetic order, Nature Communications **14**, 4145 (2023).

[36] S. Manna, A. Kamlapure, L. Cornils, T. Hänke, E. Hedegaard, M. Bremholm, B. Iversen, P. Hofmann, J. Wiebe, and R. Wiesendanger, Interfacial superconductivity in a bi-collinear antiferromagnetically ordered FeTe monolayer on a topological insulator, Nature Communications **8**, 14074 (2017).

[37] G. Chen, A. Aishwarya, M. R. Hirsbrunner, J. O. Rodriguez, L. Jiao, L. Dong, N. Mason, D. Van Harlingen, J. Harter, S. D. Wilson, *et al.*, Evidence for a robust sign-changing s-wave order parameter in monolayer films of superconducting $Fe(Se, Te)/Bi_2Te_3$, npj Quantum Materials **7**, 110 (2022).

[38] J. Checkelsky, R. Yoshimi, A. Tsukazaki, K. Takahashi, Y. Kozuka, J. Falson, M. Kawasaki, and Y. Tokura, Trajectory of the anomalous hall effect towards the quantized state in a ferromagnetic topological insulator, Nature Physics **10**, 731 (2014).

[39] M. Mogi, R. Yoshimi, A. Tsukazaki, K. Yasuda, Y. Kozuka, K. Takahashi, M. Kawasaki, and Y. Tokura, Magnetic modulation doping in topological insulators toward higher-temperature quantum anomalous hall effect, Applied Physics Letters **107**, 182401(2015).

[40] J. Wang and B. Lian, Multiple chiral Majorana fermion modes and quantum transport, Physical Review Letters **121**, 256801 (2018).

[41] J. J. He, Y. Tanaka, and N. Nagaosa, Optical responses of chiral Majorana edge states in two-dimensional topological superconductors, Physical Review Letters **126**, 237002 (2021).

[42] Hemian Yi, Yi-Fan Zhao, Ying-Ting Chan, Jiaqi Cai, Ruobing Mei, Xianxin Wu, Zi-Jie Yan, Ling-Jie Zhou, Ruoxi Zhang, Zihao Wang *et al.*, Interface-induced superconductivity in magnetic topological insulators, Science **383**, 634 (2024).

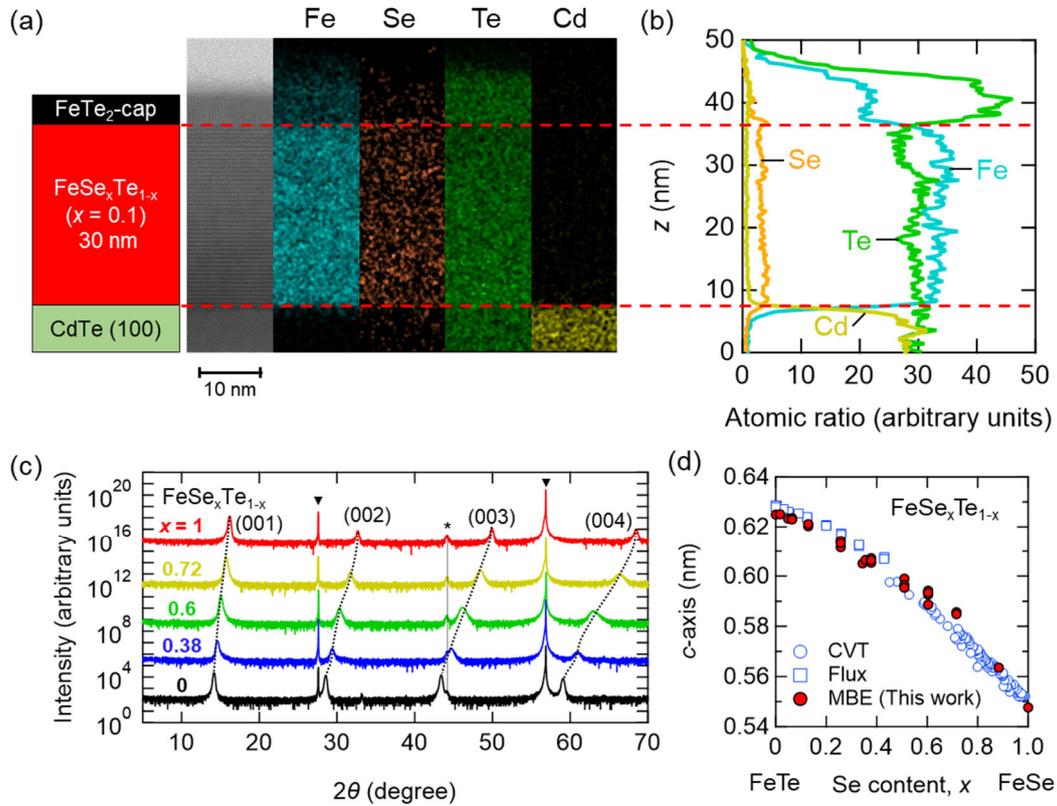

FIG. 1. **Structural characterization of FeSe$_x$Te$_{1-x}$ thin films.** (a) Real space characterization of FeSe$_x$Te$_{1-x}$ (FST) ($x = 0.1$) thin films. (Left) A schematic of FST thin film. (Middle) Cross-sectional high-angle annular dark-field scanning transmission electron microscopy (HAADF-STEM) image. (Right four panels) Elemental mappings of comprising elements of Fe, Se, Te, and Cd studied by an energy dispersive X-ray spectroscopy (EDX) for the identical area shown in the HAADF-STEM image. The red dashed line depicts the boundary between each layer. (b) Laterally integrated atomic ratio profiles obtained by the EDX in (a). The vertical axis denotes the distance along out-of-film direction $z$, whose scale is the same as (a). (c) Semi-log plots for X-ray diffraction (XRD) data of FST thin films ($x = 0$, 0.38, 0.6, 0.72, and 1). Each data is vertically shifted for clarity. The dotted lines are guides for eyes, tracing the shift of main peaks from FST(00$n$). The $x$-independent peaks (down-pointing triangle and asterisk) are from CdTe substrate. (d) Lattice constant $c$ of FST films with $t = 40$ nm as a function of Se content $x$. The open blue circle and square are reference data obtained in bulk single crystals which are grown by chemical vapor transport [24] and flux method [17], respectively.

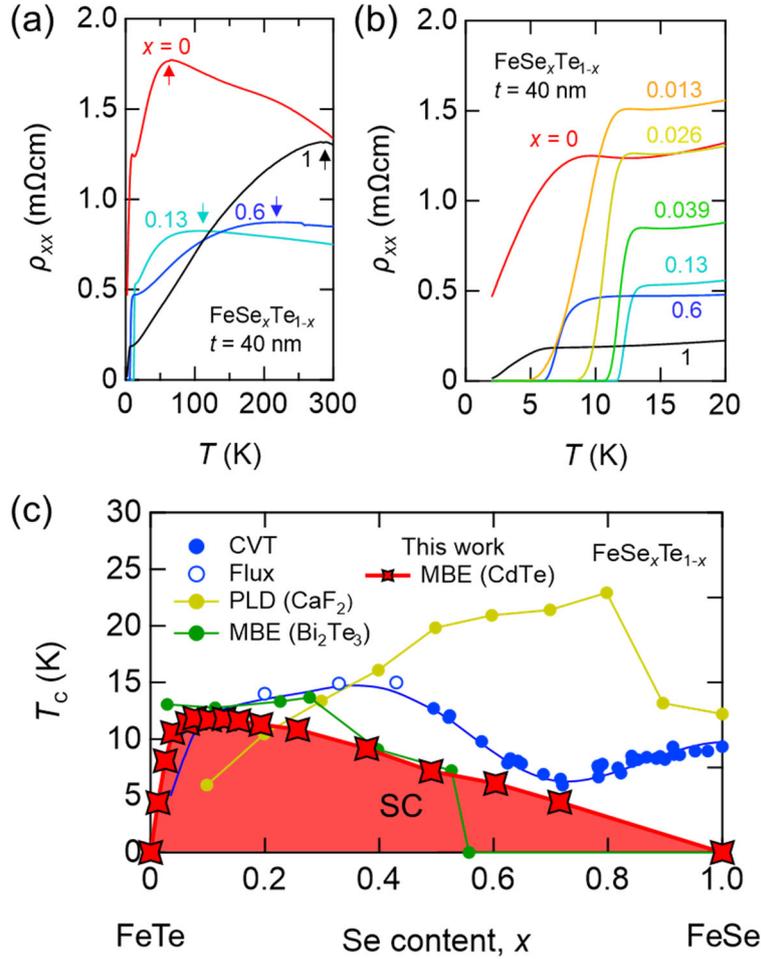

FIG. 2. **Superconductivity in FeSe$_x$Te$_{1-x}$ (FST) thin films grown on CdTe substrate.** (a) Resistivity–temperature ($\rho_{xx}$-$T$) curves for FST thin films with different Se content $x$. The arrows indicate the temperature where $\rho_{xx}$ takes its maximum. (b) The magnified $\rho_{xx}$-$T$ curves below 20 K. (c) Phase diagram of FST as a function of $x$. The superconducting transition temperature $T_c$ is defined by the zero resistance. $T_c$ and Superconducting region for FST thin films with $t$ = 40 nm are represented as red stars and red hatched area, respectively. The other markers represent $T_c$ obtained by bulk single crystals grown by chemical vapor transport (CVT) (blue filled circle [24]) and flux method (blue open circle [17]), thin films grown by pulsed laser deposition (PLD) on CaF$_2$ (yellow circle [25]), and molecular beam epitaxy (MBE) on Bi$_2$Te$_3$ (green circle [26]).

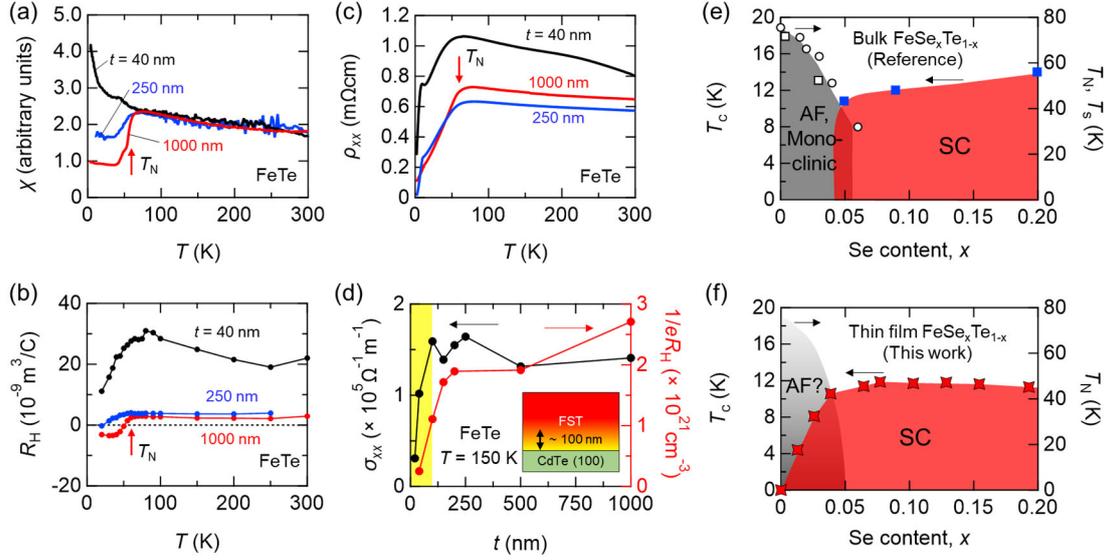

FIG. 3. **Suppressed structural phase transition and robust superconductivity in Te rich FeSe$_x$Te$_{1-x}$.** (a) Temperature $T$ dependence of magnetic susceptibility $\chi$ of FeTe films with different $t$, in which contribution from the substrate is subtracted. The red arrow indicates Néel temperature $T_N$ for the $t$ = 1000 nm sample. (b) $T$ dependence of Hall coefficient $R_H$ of FeTe films with different $t$. (c) $T$ dependence of resistivity $\rho_{xx}$ of FeTe films with different $t$. (d) $t$ dependence of conductivity $\sigma_{xx}$ and $1/eR_H$ for FeTe at 150 K. The yellow region describes the thickness region where electronic structure is significantly modified by the substrate effect. The inset depicts schematic of FST film. (e) $T$-$x$ phase diagram of bulk single crystals and (f) the present $t$ = 40 nm thin films in Te-rich region ($0 \le x \le 0.2$). The blue square [17] in (e) and red star in (f) shows superconducting transition temperature $T_c$, while white circle [18] and square [17] represent $T_N$ which coincides with the structural phase transition temperature $T_s$.

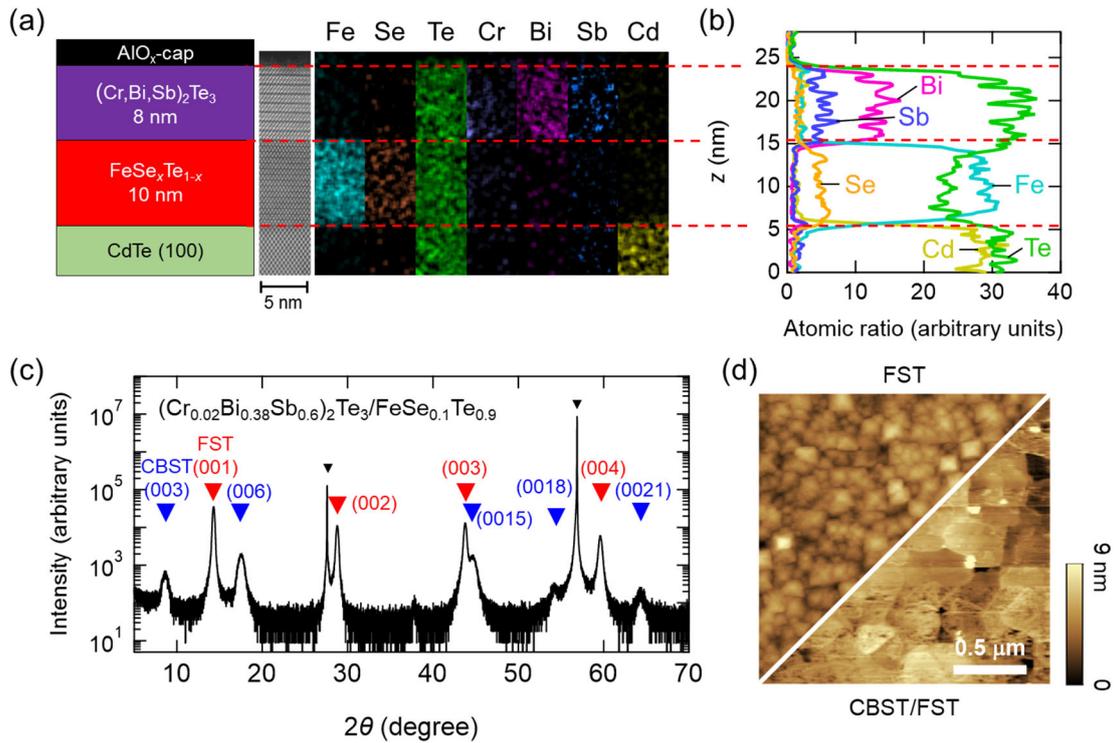

FIG. 4. **Structural characterization of $(Cr, Bi, Sb)_2Te_3$ (CBST)/$FeSe_xTe_{1-x}$ (FST) heterostructure.** (a) Real space characterization of CBST/FST ($x = 0.1$). (Left) A schematic of the sample. (Middle) Cross-sectional high angle annular dark-field scanning transmission electron microscopy (HAADF-STEM) image. (Right seven panels) Elemental mappings of comprising elements of Fe, Se, Te, Cr, Bi, Sb, and Cd obtained by an energy dispersive X-ray spectroscopy (EDX) for the same area shown in the HAADF-STEM image. The red dashed line depicts the boundary between the respective layers. (b) Laterally integrated atomic ratio profiles obtained by the EDX study in (a). (c) Semi-log plot for X-ray diffraction (XRD) data of CBST/FST ($x = 0.1$). The blue, red, and black down-pointing triangles denote the peaks from CBST($003n$), FST($00m$), and CdTe($l00$), respectively. (d) Atomic force microscopic image of FST (top left) and CBST/FST (bottom right). The white scale bar represents 0.5 $\mu$m.

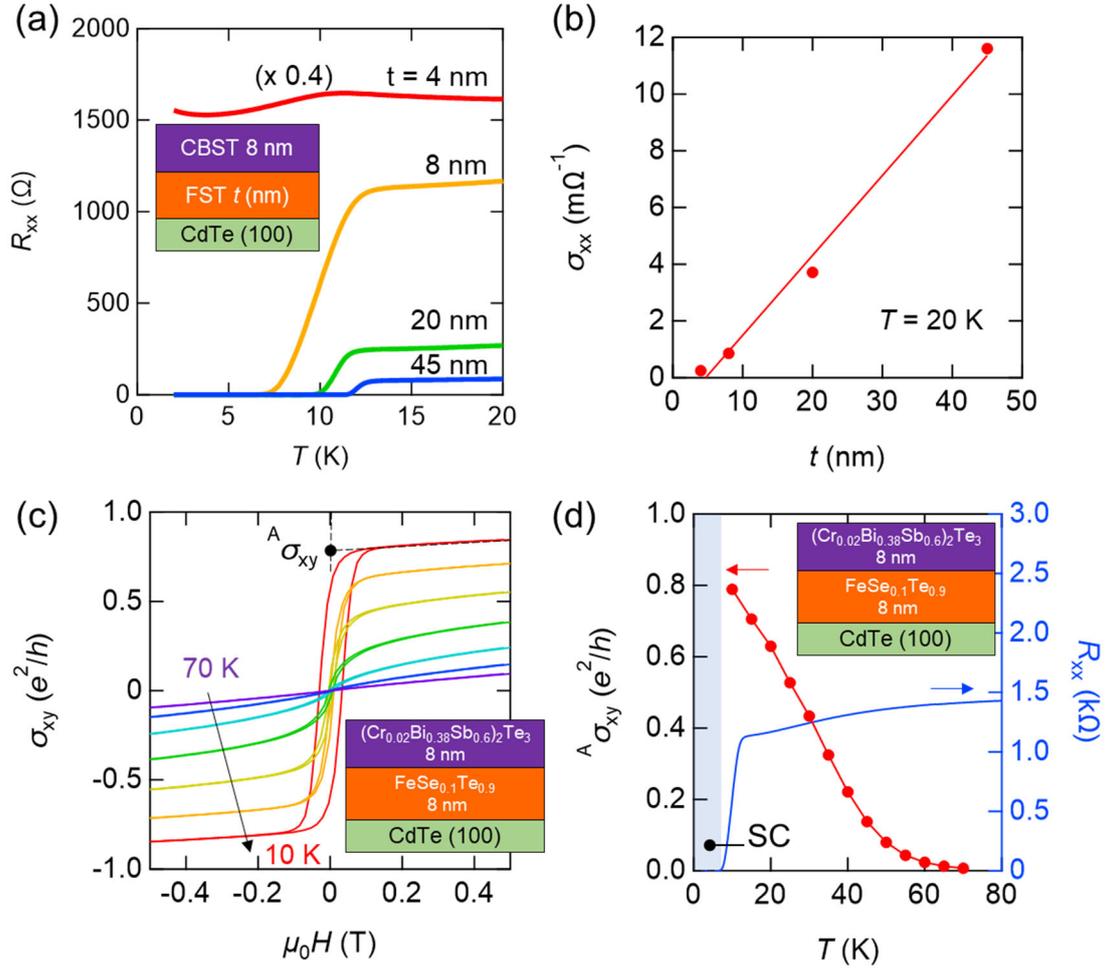

FIG. 5. **Superconductivity and large anomalous Hall effect in a magnetic topological insulator/ superconductor heterostructure.** (a) Temperature $T$ dependence of longitudinal sheet resistance $R_{xx}$ of $(Cr_{0.02}Bi_{0.38}Sb_{0.6})_2Te_3$ (CBST)/FeSe$_x$Te$_{1-x}$ (FST) ($x = 0.1$) heterostructure with different thickness $t$ of FST. The inset is schematics of our device. (b) Sheet conductivity $\sigma_{xx}$ vs. $t$ at 20 K. The line is a linear fit for the data. (c) Field $H$ dependence of sheet Hall conductivity $\sigma_{xy}$ of CBST/FST ($x = 0.1$ and $t = 8$ nm). The data are taken from 10 K to 70 K with 10 K intervals. Anomalous Hall conductivity $^A\sigma_{xy}$ is determined by the extrapolation for each data from high field. (d) $^A\sigma_{xy}$ (left axis) and $R_{xx}$ (right axis) as a function of $T$. The shaded area is superconducting phase, where $R_{xx}$ drops to zero and Hall conductivity is not measurable.

# Supplemental Materials: Molecular beam epitaxy of superconducting FeSe$_x$Te$_{1-x}$ thin films interfaced with magnetic topological insulators

## I.   DETERMINATION OF Se CONTENT BY INDUCTIVELY COUPLED PLASMA MASS SPECTROMETRY

To determine the actual Se content in the FeSe$_x$Te$_{1-x}$ (FST) thin films, we performed inductively coupled plasma mass spectrometry (ICP-MS) analysis on several FST samples with different nominal flux ratios $x_n$. To prevent overestimation of Te elements, which are also present in the CdTe substrate, we in turn used LaAlO$_3$(100) substrate only for this experiment. The result is shown in Fig. S1, where $x$ determined by ICP-MS is plotted against $x_n$. We found $x$ is nearly proportional to $x_n$, which indicates that the actual Se content in the FST thin films is almost the same as nominal equivalent beam pressure ratio of Se and Te. We note that this relationship is advantageous in controlling $x$. The values of $x$ used in the main text is calibrated by interpolating between the nearest two points in Fig. S1.

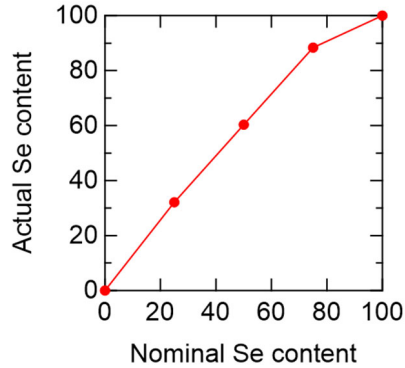

FIG. S1. **Actual Se content $x$ determined by inductively coupled plasma mass spectrometry as a function of nominal $x_n$.**

## II.   RECIPROCAL SPACE MAPPINGS

Evaluating in-plane strain is crucial for discussing the electronic properties of FST thin films. To evaluate strains, we conducted reciprocal space mapping measurements for FeTe films with different thickness $t$. The sharp asymmetric Bragg peak of FeTe(104) is clearly resolved for the thick FeTe film with $t$ = 1000 nm, as shown in Fig. S2(a). A Bragg peak expected for bulk single crystals of FeTe is depicted as the cyan markers. We observed that the peak center aligns well with the literature along the out-of-plane $Q_z$ direction, while it broadens along the in-plane $Q_x$ direction. It is noteworthy that the observed Bragg peak slightly shifts to lower $Q_x$ compared to the bulk, indicating the presence of tensile strain. As $t$ decreases, the intensity of the Bragg peak significantly decreases, and it becomes faint for the $t = 40$ nm sample, as shown in Figs. S2(b)-(d). This is likely due to incoherence of the lattice, especially along the in-plane direction for thinner films. The peak position seems to consistently shift to lower $Q_x$, as demonstrated by the Gaussian fittings in Figs. S2(e)-(h). In-plane lattice constant $a$ is plotted as a function of $t$, in Figs. S3(a)-(b). We observed a trend where $a$ becomes longer with $t$ decreases, suggesting the stronger tensile strain in thinner films. The tensile strain for the 100 nm sample is evaluated to be $\Delta a/a \sim 0.4\%$ compared to a bulk. We note that the same order of tensile strain is reported in superconducting FeTe thin films [1]. Additionally, as presented in Fig. 1(d) of the main text, the value of $c$ for a FeTe film with $t = 40$ nm is slightly smaller than that of bulk crystals, supporting the presence of the in-plane tensile strain. These observations suggest that in-plane tensile strain remains significant in FeTe films with $t < 100$ nm. We speculate that this in-plane tensile strain could be one of the factors that alter the electronic structures of FST thin films.

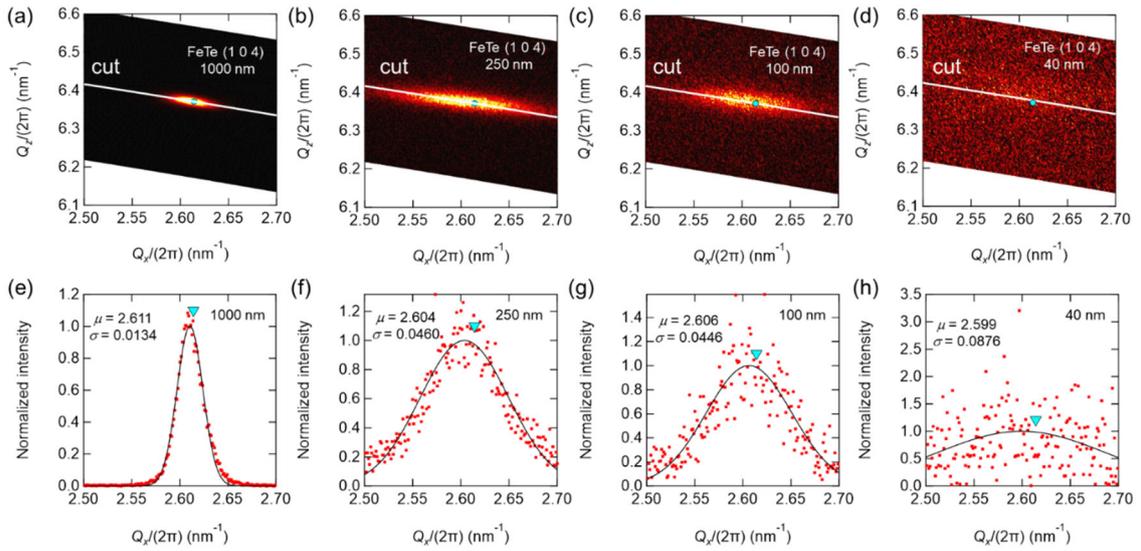

Fig. S2 **Reciprocal space mappings.** The measurements were conducted for FeTe films with $t =$ (a) 1000 nm, (b) 250 nm, (c) 100 nm, and (d) 40 nm, near the (104) Bragg peak. The cyan markers denote the Bragg peak expected for bulk single crystals of FeTe. (e)-(h) Gaussian fittings along the line cut presented in (a)-(d), respectively.

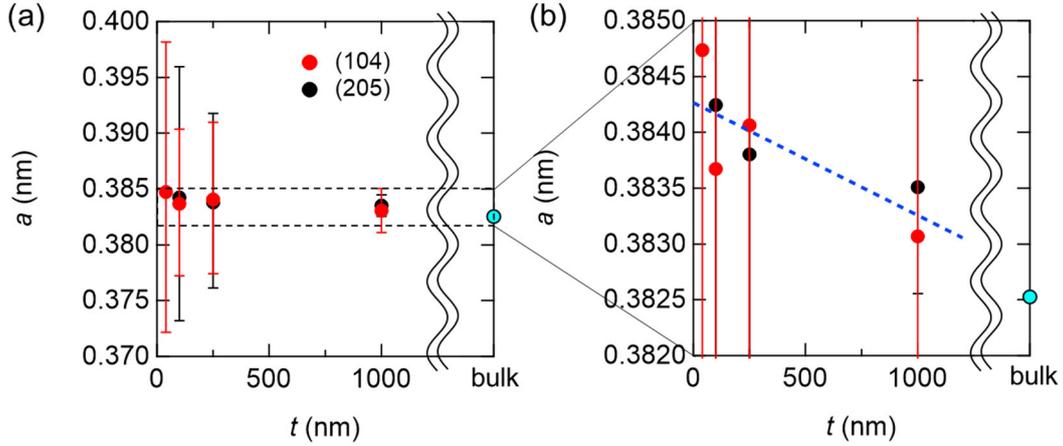

Fig. S3 **In-plane lattice constant *a* of FeTe as a function of thickness *t*.** The red and black markers are obtained by the (104) and (205) Bragg reflections, respectively. The error bars are calculated by standard deviations of the gaussian fittings. (b) The magnified view of the dashed box in (a). The blue dashed line is a fit to the data.

## III. Te-ANNEALING EFFECTS

Over the past decades, it has been widely recognized that superconductivity in the FST family is very sensitive to perturbation in stoichiometry. As-grown FST tends to contain excess Fe, which plays a role for magnetic impurity scattering target and dramatically suppresses superconductivity. It has been reported that many types of annealing procedures can remove the excess iron impurity and enhance $T_c$ by, for example, annealing the sample in Te atmosphere [2]. To examine how Te-annealing improves transport properties in the MBE-grown FST, we characterized sets of FST films with different annealing conditions. The growth recipe is given in Fig. S4(a), where we grow FST at 240°C and the sample is cooled in a Te atmosphere for the duration of $t_A$. We take out the sample when the substrate temperature reaches 100°C, typically 60 minutes after the growth. We characterized the FST films with different Te-annealing time $t_A$ using XRD, as shown in Fig. S4(b). We observe that peak position and width of FST(001) are little affected by changes in $t_A$, indicating that the films keeps the same composition $x$ and homogeneity during the Te-annealing. We extract the peak position and convert it to the lattice constant $c$ (Fig. S4(c)). All films exhibit almost the same $c$ within the error denoted by the shaded area in Fig. S4(c). Assuming linear relation of $c \propto x$, the error in $x$, propagated by that in $c$, corresponds to $\Delta x \sim \pm 1$ %. This small error in $c$ may result from fluctuations in flux pressures for the Se and Te sources, which can vary by about ±1 % during the growth. Therefore, one cannot conclude whether the error in $c$ arises from the fluctuations of the flux pressures or from the Te-annealing procedure.

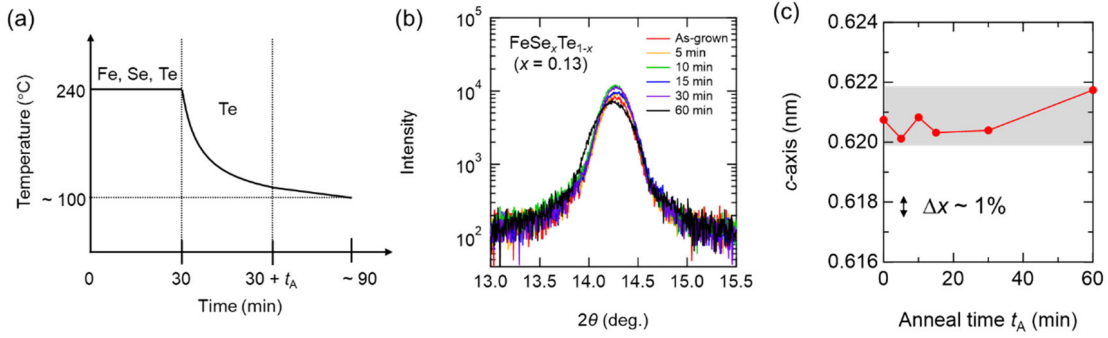

FIG. S4. **Te-annealing and its effect on crystal structure.** (a) Growth recipe for FST thin films. We keep supplying Te flux during the cool-down process for time of $t_A$. (b) XRD data for FST thin films with different $t_A$. (c) $c$-axis lattice constant as a function of $t_A$. The shaded area shows error range, which corresponds to $x \sim \pm 1$ %.

While Te-annealing has minimal effect on the crystal structure, electronic transport properties can be dramatically influenced. We present temperature dependence of resistivity for nearly optimal composition FST ($x = 0.13$) films with different $t_A$ up to 60 minutes in Figs. S5(a) and S5(b). As-grown FST shows insulating behavior from room temperature to the superconducting transition, and $\rho_{xx}$ reaches to zero at around 4.5 K. With increasing $t_A$, the insulating behavior gradually diminishes, and the system becomes metallic below a coherent crossover temperature of $\sim$ 70 K. $T_c$ monotonically increase as well and saturates to its optimal value of 12 K above $t_A = 30$ minutes. The transport data presented in the main text are obtained by well-annealed samples, where $T_c$ is optimized with respect to $t_A$.

Even though the Te-annealing does not affect the peak center in XRD, it is still possible that only the top few layers would have a larger Te content. If these Te-rich few layers exhibit a higher $T_c$ than the rest of the FST layers, it could potentially "improve" $T_c$, in the phase diagram. Let us suppose that the Te-rich top layers have a lower Se content $x$-$\delta$ ($\delta > 0$), while the rest of the FST has $x$, as illustrated in Fig. S6(a). In Fig. S6(b), we depict a hypothetical phase diagram of FST for this sample. In the underdoped region ($x < x_0$), the Te-rich top layers have a lower Tc compared to the rest of FST, resulting in an unchanged $T_c$. On the other hand, in the overdoped region ($x > x_0$), the top Te-rich FST would have a higher $T_c$ than that of the interior, resulting in an enhanced $T_c$. Therefore, the effect of the Te-rich layers could only shift the superconducting dome to higher $x$. Importantly, the optimal $T_c$ for the Te-annealed FST ($\sim$ 12 K) is significantly larger than that of the as-grown FST films ($\sim$ 7.5 K), as shown in Fig. S6(c). This indicates that the Te-annealing process has a Tc enhancement effect. Additionally, the variation in Se content $\delta$ would also result in a plateau-like region near the optimal doping $x_0 < x < x_0+\delta$. Indeed, we find that there is a certain doping range of $0.1 < x < 0.2$ where $T_c$ remains nearly optimal. This indicates that the upper limit for $\delta$ should be less than 0.1. Therefore, in the worst-case scenario of $\delta$ = 0.1, the true $T_c(x)$ might look like the dashed blue line presented in Fig. S6(c). However, the main results of the present study, i.e., optimal $T_c$ = 12 K at around $x$ = 0.13, and the robust superconductivity in the underdoped region $x$ < 0.1, would not be affected by the presence of the possible Te-rich top few layers.

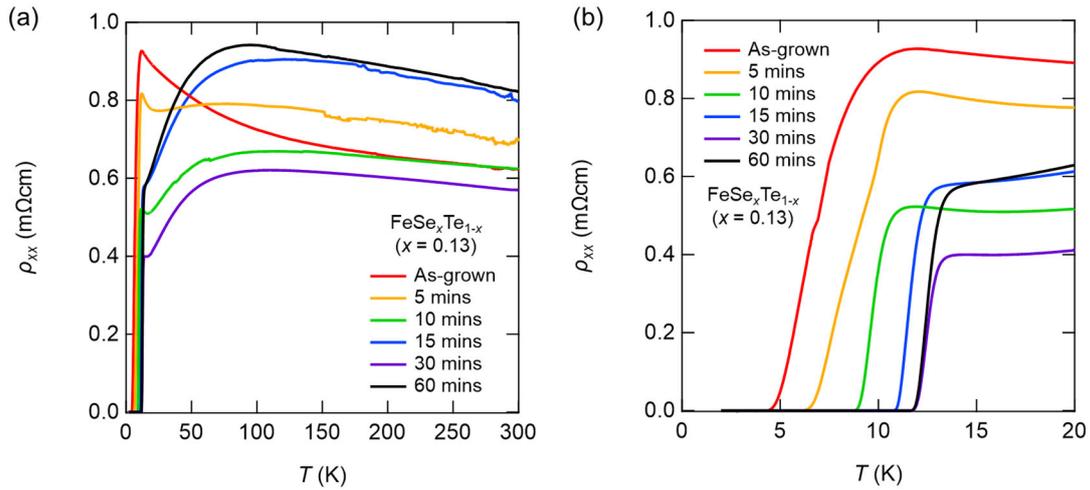

FIG. S5. **Temperature dependence of resistivity $\rho_{xx}$ with varying Te-annealing time.** The same data is plotted (a) below 300 K and (b) below 20 K.

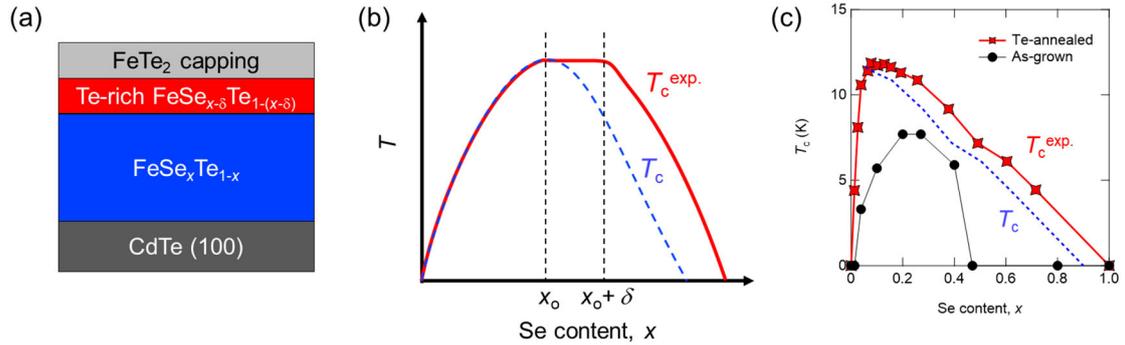

FIG. S6. **Effect of possible Te-rich top layers formed during the Te-annealing.** (a) Schematic of a FST thin film where the top few layers have a lower Se content $x$-$\delta$. (b) Schematics of the true (the blue dashed line) and experimentally observable (the red line) $T_c$-$x$ phase diagrams. (c) $T_c$-$x$ phase diagrams for the Te-annealed FST and as-grown FST thin films. The blue dashed line represents $T_c$ in the case of $\delta$ = 0.1.

## IV.    FORMATION OF FeTe$_2$ CAPPING LAYER

A long-duration Te-annealing process can result in the formation of an orthorhombic marcasite FeTe$_2$ capping layer, typically 10 nm thick. This phase tends to prefer relatively lower growth temperatures and is deposited on top of FST thin films. We present XRD data for a Te-annealed FeTe thin film in Fig. S7(a). In addition to the peaks from FeTe and CdTe substrate, a minor peak has been found near $2\theta \sim 33°$, which matches well with the peak position expected for FeTe$_2$(200). The presence of the FeTe$_2$ phase is also supported by cross-sectional high angle annular dark-field scanning transmission electron microscopy (HAADF-STEM) image, as shown in Fig. S7(b). At the bottom of the picture, we observed atomically resolved FST ($x = 0.1$) layers which stack along the out-of-plane direction. On top of the FST layer, a periodic pattern of vertical faint stripes can be resolved. We attribute this pattern to the FeTe$_2$ impurity phase, where the period along the horizontal direction roughly agrees with the $c$-axis length of FeTe$_2$. XRD peaks from FeTe$_2$ can be only found for the FeTe$_2$(200), suggesting that the $a$-axis of FeTe$_2$ is oriented toward the film's out-of-plane direction, which is consistent with the HAADF-STEM observation. It has been reported that FeTe$_2$ is an insulating magnet which exhibits antiferromagnetism and ferromagnetism at around 80 K and 35 K, respectively [3]. The resistivity of FeTe$_2$ $\rho_{FeTe2}$ is about 10 times larger than that of FST $\rho_{FST}$ at 300 K, and this ratio $\rho_{FeTe2}/\rho_{FST}$ further increases as the temperature decreases. Near 20 K, $\rho_{FeTe2}/\rho_{FST}$ reaches more than 300. Any anomalies associated with the magnetic transitions of FeTe$_2$ cannot be observed in any FST thin films, indicating little contribution of FeTe$_2$ to the electrical transport. Regarding the magnetization measurements, on the other hand, the influence of FeTe$_2$ may not be negligible especially for the thinnest FeTe film ($t = 40$ nm), where the volume fraction of FeTe$_2$ becomes comparable to that of FeTe. Indeed, we observed an upturn in the magnetic susceptibility of the 40 nm FeTe thin film below about 40 K (Fig. 2(a) in the main text), which may be attributed to the ferromagnetic transition of FeTe$_2$. Nevertheless, no discernible drop in $\chi$ associated with the AF transition of FeTe ($T_N \sim 60$ K) is resolved, confirming the suppressed tetragonal to monoclinic structural transition. Therefore, the presence of the FeTe$_2$ layer does not affect our conclusion.

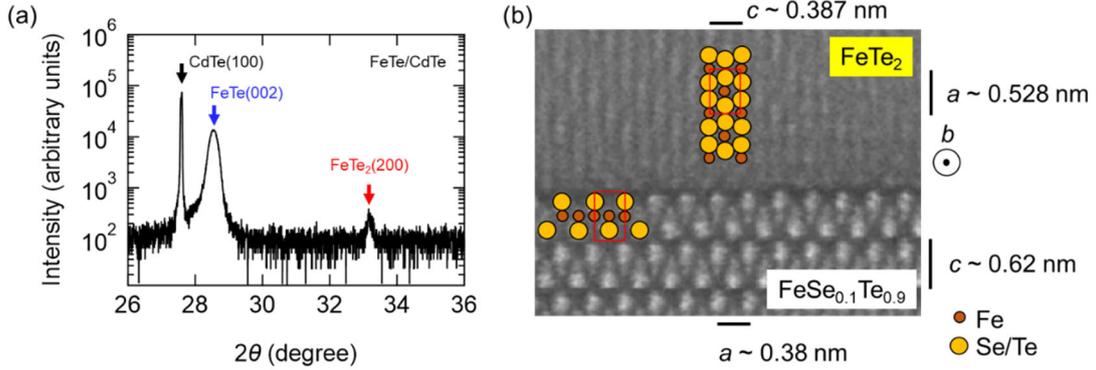

FIG. S7. **Formation of FeTe₂ capping layer during the Te-annealing process.** (a) XRD data for FeTe thin film after Te-annealing for 60 minutes. The minor peak around $2\theta \sim 33°$ can be attributed to $FeTe_2(200)$. (b) Cross-sectional high angle annular dark-field scanning transmission electron microscopy (HAADF-STEM) image near an interface between $FeSe_{0.1}Te_{0.9}$ and $FeTe_2$. The brown and yellow circles denote Fe and Se/Te irons, respectively. The red rectangles indicate unit cells for each compound.

## V.    SUPERCONDUCTIVITY IN THE Te RICH FST THIN FILMS

We present transport data in Te-rich FST in more detail in Fig. S8(a). We find that even a few percentages of Se inclusion can dramatically improve $T_c$. Zero resistivity is clearly observed for all the samples except $x = 0$. We note that zero resistivity has never reported in bulk crystals of FST with $x \leq 0.05$, where the AFM ordering and accompanying structural transition become more stable. For the end compound FeTe, the onset of superconducting transition is observed in samples with $t = 40, 150$, and 1000 nm, as shown in Fig. S8(b). The onset temperature $T_c^{onset}$ shows a slight increase with increasing $t$ (Fig. S8(c)). This behavior is likely attributed to the suppression of disorder, as previously reported in thin films of FeSe [4].

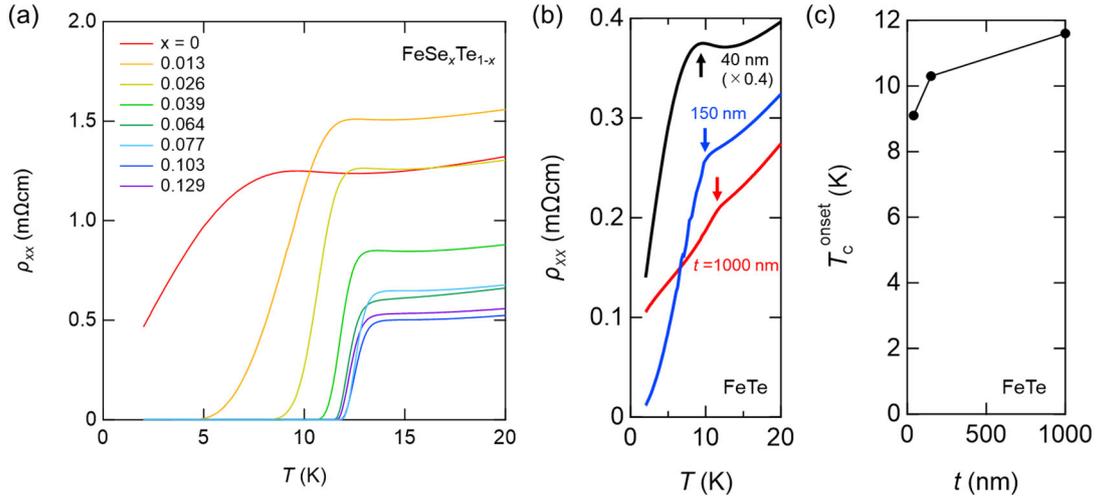

FIG. S8. **Superconductivity in the Te-rich FST films.** (a) $\rho_{xx}$-$T$ curves of Te-rich FST thin films ($0 \leq x \leq 0.129$). (b) $\rho_{xx}$-$T$ curves of FeTe films with different $t$. The arrows represent $T_c^{\text{onset}}$. (c) $t$-dependence of $T_c^{\text{onset}}$ for FeTe films.

## VI.    HALL EFFECT OF FST THIN FILMS

Measurement of the Hall coefficient can provide pivotal information on the fermiology of a material, including carrier type, carrier density, and mobility. In iron-based superconductors, multiple Fe $3d$ orbitals appear near the Fermi level, usually hosting multi-band characteristics. This two-band character can be extracted from the non-linear behavior of the Hall resistance $R_{yx}$ vs. magnetic field $H$ [5]. We present representative data for $R_{yx}$ in FeTe thin films with $t = 40$ nm and 1000 nm in Fig. S9(a) and S9(b), respectively. All the $R_{yx}$ data show nearly linear in $H$ behavior up to 14 T at all measurement temperatures. It is, therefore, difficult to separately estimate carrier concentrations and mobilities for electron and hole pockets from the current sets of data. Further studies of high field experiments are needed to explore the fermiology in FST thin films.

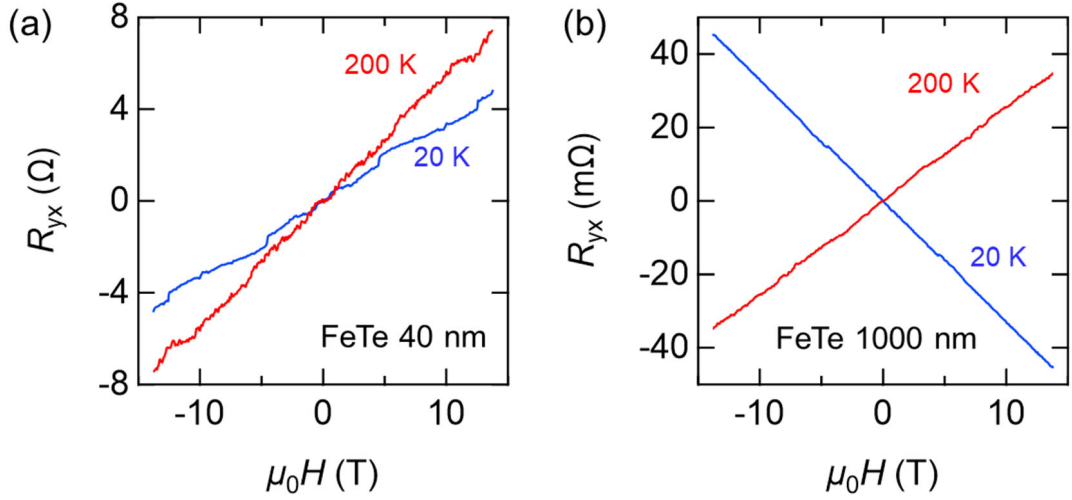

FIG. S9. **Field dependence of Hall resistance for FeTe films.** The data is taken for FeTe films with thickness of (a) 40 nm and (b) 1000 nm, respectively.

## VII.    MAGNETIC SUSCEPTIBILITY MEASUREMENTS

We performed magnetic susceptibility measurements using Magnetic Properties Measurement System (Quantum design) for the MBE grown FST thin films. We first measured the magnetic response from a bare CdTe substrate. In Fig. S10(a), the $H$-dependence of magnetization $M$ of CdTe at 2 K is depicted. We observed linear $M$-$H$ data with a negative slope, indicating diamagnetic response from CdTe. Subsequently, we performed a temperature dependent susceptibility, $\chi = M/(\mu_0 H)$ scan at 7 T, as shown in Fig. S10(b). We observed that the diamagnetic response is weakly $T$ dependent, which is favorable when we measure small $M$ from the tiny volume of the FST thin film on top of the CdTe substrate. We then measured the magnetization of FST thin films and subtracted the contribution from CdTe. We note that reliable estimation of the absolute value of $\chi$ for FST is challenging because of the small signal (FST) to background (CdTe) ratio, which is inevitable for our experimental setup. However, qualitative behavior of $\chi$ is consistent with that of FST bulk single crystals, especially when the sample is thick enough (for example, $t = 1000$ nm). In Fig. S11, we show $\chi$-$T$ curves at 7 T for thick FST films ($t = 1000$ nm) with different $x$. At higher temperature, all the samples show paramagnetic behavior with a similar slope. A relatively sharp drop of $\chi$ is observed, especially for $x = 0$ and 0.026, where AFM ordering is expected below $T_N$. For samples with $x \geq 0.064$, where no AFM ordering is reported in bulk single crystals [2], the drop is still visible but becomes broader. Furthermore, $\chi$ for zero field cooling (ZFC) decreases below 10-20 K, while the reduction is not significant for the field cooling (FC) process. This anomaly could be attributed to the superconducting diamagnetic response.

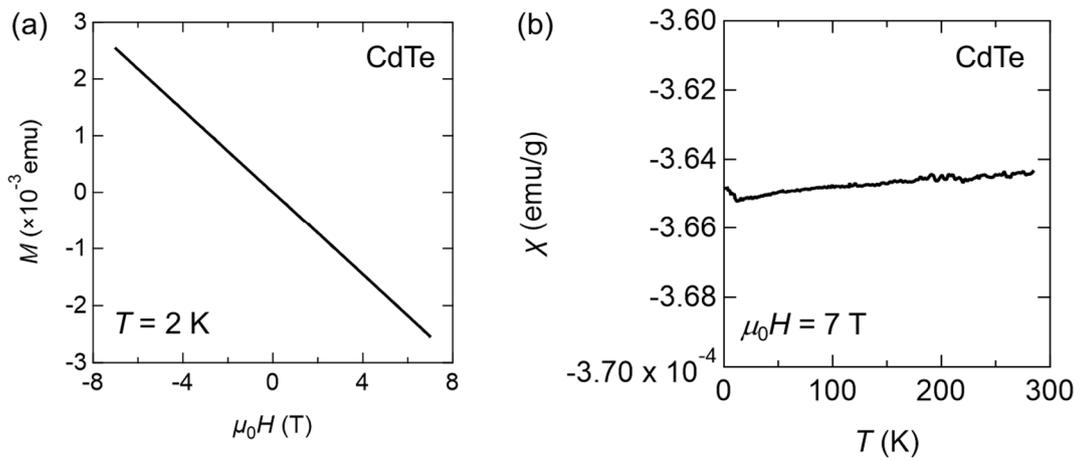

FIG. S10. **Magnetization measurement of CdTe substrate.** (a) Field $H$ dependence of magnetization $M$ at 2 K. (a) Temperature $T$ dependence of magnetic susceptibility $\chi$ at 7 T.

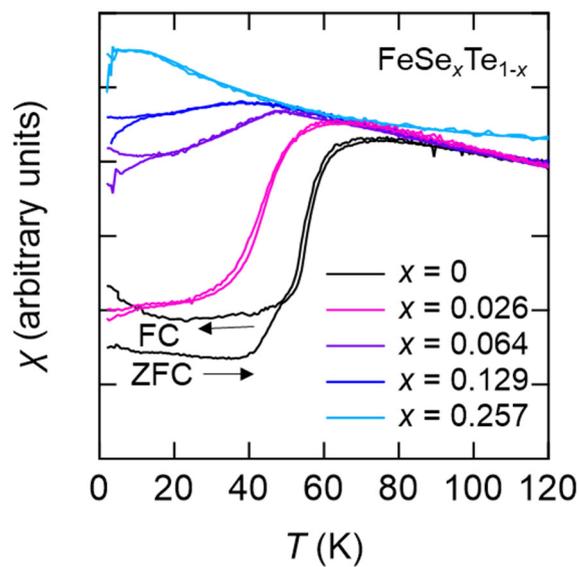

FIG. S11. **Temperature dependence of susceptibility of FST thick films.** Thickness for all the samples is 1000 nm.

## VIII.   TRANSPORT PROPERTIES OF THE TOPOLOGICAL INSULATOR/SUPERCONDUCTOR HETEROSTRUCTURES

We depict the temperature dependence of sheet resistance for $Bi_2Te_3(BT)/FeSe_{0.1}Te_{0.9}$ heterostructures in Fig. S12(a) with different thickness $t$ for the FST layer. The optimal $T_c = 12$ K is achieved for $t = 45$ nm, and as $t$ decreases, $T_c$ is gradually reduced. This trend is the same as that of CBST/FST heterostructures, which is presented in the main text (Fig. 5(a)). While superconductivity vanishes for the 4 nm CBST/FST, it is present for 4 nm BT/FST. The difference then should be mainly attributed to the magnetic impurity of Cr, which can partially suppress superconductivity near the interface. We also show the $H$-dependence of sheet Hall resistance $R_{yx}$ of an FST thin film and CBST/FST heterostructures at 15 K, as presented in Fig. S12(b). The FST film does not show an anomalous Hall effect but shows an ordinary Hall effect with a positive slope. On the other hand, the CBST/FST heterostructure show a distinct anomalous Hall effect, and interestingly, it also shows a positive ordinary Hall component with nearly the same amplitude as that of the FST single layer. Furthermore, this Hall coefficient is nearly independent of Cr concentration, although the amplitude of the anomalous Hall component and coercive field highly depend on it. The results indicate that the observed positive ordinary Hall coefficient mainly comes from FST, whereas CBST shows a small contribution to it.

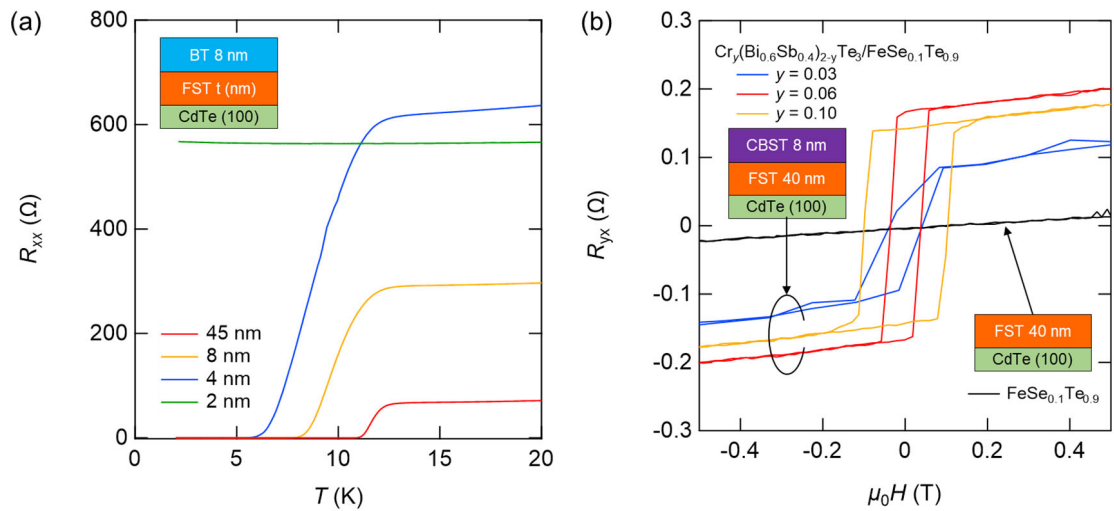

FIG. S12. **Transport properties of topological superconductor/ topological insulator heterostructures.**
(a) Sheet resistance $R_{xx}$ as a function of $T$ for non-magnetic heterostructure of BT/FST with different thickness of

FST. (b) Field dependence of sheet Hall resistance $R_{yx}$ for FST single layer, and CBST/FTS heterostructures. The latter one includes three samples with different Cr concentrations.

---

[1] Y. Han *et al.*, Superconductivity in Iron Telluride Thin Films under Tensile Stress, Physical Review Letters, **104**, 017003 (2010).

[2] Y. Sun, Z. Shi, and T. Tamegai, Review of annealing effects and superconductivity in $Fe_{1+y}Te_{1-x}Se_x$ superconductors, Superconductor Science and Technology **32**, 103001 (2019).

[3] A. Rahman *et al.*, Multiple magnetic phase transitions, electrical and optical properties of $FeTe_2$ single crystals, Journal of Physics: Condensed Matter **32**, 035808 (2020).

[4] R. Schneider *et al.*, Superconductor-insulator quantum phase transition in disordered FeSe thin films, Physical Review Letters, **108**, 257003 (2012).

[5] M. Čulo *et al.*, Putative Hall response of the strange metal component in $FeSe_{1-x}S_x$, Physical Review Research **3**, 023069 (2021).